\documentclass[twocolumn,aps,prb]{revtex4-2}
\usepackage{graphicx,adjustbox,multirow,latexsym,amssymb,amsmath,bbold,paralist,braket,appendix}

\newcommand{\cphase}{$Fm\bar{3}m$}
\newcommand{\tphase}{$P4_{2}nmc$}
\newcommand{\oI}{$Pca2_{1}$}
\newcommand{\oII}{$Pmn2_{1}$}
\newcommand{\mHfo}{$P2_{1}/c$}
\newcommand{\aep}{a$_{\text{epi}}$}

\newcommand{\HZO}{Hf$_{0.5}$Zr$_{0.5}$O$_{2}$}

\begin{document}

\title{Hafnia HfO$_2$ is a Proper Ferroelectric}

\author{Aldo Raeliarijaona}
\email{araeliarijaona@carnegiescience.edu}
\author{R. E. Cohen}
\email{rcohen@carnegiescience.edu}
\affiliation{Extreme Materials Initiative, Earth and Planets Laboratory, Carnegie Institution for Science, 5241 Broad Branch Road NW, Washington, DC 20015, USA}

\date{\today}

\begin{abstract}
We clarify the nature of hafnia as a proper ferroelectric and show that there is a shallow double well involving a single soft polar mode as in well-known classic ferroelectrics. Using symmetry analysis, density-functional theory (DFT) structural optimizations with and without epitaxial strain, and density functional perturbation theory (DFPT), we examine several important possible hafnia structures derived ultimately from the cubic fluorite structure, including  baddeleyite (\mHfo), tetragonal antiferroelectric \tphase, $Pbca$ (nonpolar and brookite), ferroelectric rhombohedral ($R3m$ and $R3$), \oII, and \oI~structures. The latter is considered to be the most likely ferroelectric phase seen experimentally, and has an antiferroelectric parent with space group $Pbcn$, with a single unstable polar mode and a shallow double well with a well depth of 24 meV/atom. Strain is not required for switching or other ferroelectric properties, nor is coupling of the soft-mode with any other modes within the ferroelectric \oI,\oII, $R3m$ or $R3$ phases.
\end{abstract}                      
\maketitle
\section*{Introduction}
Hafnia has great potential as a technologically important material due to  experimentally observed ferroelectricity in thin films\cite{Boscke2011,BosckeFerroelectricityIH2011} and its compatibility with silicon for use in integrated electronics\cite{Robertson2006,Hubbard_schlom1996}. The nature of the observed ferroelectricity is complicated by a complex set of potential phases and distortions from the parent cubic fluorite structure, and by the dependence of its properties and phase relations due to doping, strain, and applied electric fields that give rise to the so-called wake-up effect. Here we clarify that ferroelectric hafnia can be understood as a simple, proper, ferroelectric, from which perturbations from doping, strain, and applied fields can be explored straightforwardly.

Complications in understanding hafnia stem from the fact that there are no known stable ground state polar phases, even varying pressure and temperature \cite{Ohtaka2001}. At ambient pressure, the phase diagram shows the fluorite cubic phase (\cphase) above 2600K, tetragonal phase (\tphase) between 1800K and 2600K, and monoclinic baddeleyite (\mHfo) below 1800 K\cite{Ohtaka2001}. Increasing pressure at room temperature gives non-polar orthorhombic $Pbca$ for 5 GPa$ \leq p \leq$ 15 GPa, and Pnma for $p >$ 15 GPa.

It is often mistakenly said that hafnia is an improper ferroelectric \cite{Lee2020,Delodovici2021,Zhou2022}. This is important, because an improper ferroelectric does not have a normal ferroelectric phase transition. An improper ferroelectric, has, by definition, a non-polar primary order parameter that couples with a polar zone center mode. If the coupling between the polar mode and the non-polar mode is even, so that the polar mode can be switched without switching the sign of the coupled mode, an improper ferroelectric could still be used much as a proper ferroelectric in applications that require ferroelectric switching, but a large peak in dielectric response may not occur near the (non-ferroelectric) phase transition. 

As we show below, ferroelectric hafnia is a proper ferroelectric with a polar mode giving rise to a symmetric double well. This is true both for the orthorhombic ferroelectrics and the rhombohedral epitaxial hafnia. \cite{Wei2018}  There is no choice as to what is the proper parent structure to determine whether a ferroelectric is proper or improper. The appropriate parent structure is that of the lowest energy maximum in energy in switching the polarization by an applied electric field.  It could be a fictive or average structure if the ferroelectric has order-disorder character.

The situation is clear if we compare with well known classic perovskite ferroelectrics, which are now well understood \cite{Cohen1992,Cohen2000}. In the perovskites, the parent paraelectric phase has simple cubic perovskite structure with five atoms per unit cell, but this does not mean that the atoms all sit on their ideal sites.  Rather the atoms hop among symmetric off-centered sites, and only the average structure has the simple cubic structure. As the temperature is lowered at ambient pressure, three phase transitions are observed in BaTiO$_{3}$, for example, cubic to tetragonal ($P4mm$) around 400K, tetragonal to orthorhombic ($Amm2$) around 280K, and orthorhombic to rhombohedral ($R3m$) around 200K\cite{Kwei1993,Ishidate1997}. Although the transitions are understood to be distortions of the parent {\em average} structures due to a polar soft phonon, and only the ground state rhombohedral phase is an ordered structure with atoms vibrating around their ideal sites in the $R3m$ structure. In contrast, the classic ferroelectric LiNbO$_3$, has ten atoms per unit cell, from a zone boundary instability in the parent cubic perovskite. The resulting paraelectric R$\bar{3}$c phase has a single polar soft mode giving a proper double well to a polar, ferroelectric $R3c$ ground state. Just like in the simple perovskite ferroelectrics, the atoms do not sit on their ideal sites in the ferroelectric phase, but hop \cite{Inbar1996,Bakker1993}. Unlike BaTiO$_{3}$ etc., in LiNbO$_3$, cubic perovskite  never forms below the melting point. As we show below, hafnia is akin to LiNbO$_3$ in that the ferroelectric phase is a proper ferroelectric with a single polar mode, but unlike LiNbO$_3$, at high temperatures the parent tetragonal and cubic phases do form below the melting point, partly due to its very high melting temperature of 3031K. Indeed, similar to hafnia, originally Megaw suggested that LiNbO$_{3}$ switched through the cubic perovskite as its paraelectric parent\cite{Megaw1968}, but Barker and Loudon showed that the parent is $R\bar{3}c$\cite{Barker1967}, which was confirmed in later work \cite{Inbar1996,Scrymgeour2005}.

At high temperatures, cubic fluorite hafnia has a soft zone-boundary mode, $X^{-}_{2}$, leading to tetragonal \cite{Lee2020} at 2600K. However, tetragonal hafnia is dynamically stable with a first-order phase transition to monoclinic baddeleyite at 1800K. The absence of a polar, stable ferroelectric phase made the discovery of ferroelectricity in thin film hafnia a great surprise\cite{Boscke2011}.

Previous experiments and theory show that ferroelectric hafnia is most likely orthorhombic with space group \oI \cite{Sang2015,Shimizu2015}. The energy barriers between ferroelectric \oI, the non-polar ground state baddeleyite \mHfo, orthorhombic antiferroelectric $Pbcn$, and tetragonal antiferroelectric \tphase~are all large, larger than kT at room temperature for pure hafnia\cite{Liu2019}. However, once formed, ferroelectric \oI~is metastable and switchable. There have a been a number of proposals for formation of the ferroelectric phase, including doping and vacancies\cite{Hoffmann2015}, strain\cite{Liu2019}, kinetics\cite{Xu2021}, and electric field\cite{Batra2017,Materlik2015}. Other polar phases such as \oII~\cite{Huan2014,Qi2020} and rhombohedral $R3m$\cite{Wei2018,Kaiser2023} are also metastable albeit with energies larger than that of \oI. \oII~has been only proposed theoretically\cite{Huan2014,Qi2020}, whereas the $R3m$ has been experimentally observed in (111)-oriented \HZO\cite{Wei2018,Kaiser2023}. Ref.~\onlinecite{Nentwich2022} presented the tree of subgroups from fluorite in hafnia (and zirconia), mostly consistent with our work, though we concentrate on the important phases leading to ferroelectricity, and neglect any phases with partially occupied sites, which are improbable. We concentrate on the symmetry modes (soft and hard modes) over  group/subgroup relations. We discuss differences from Nentwich's\cite{Nentwich2022} analysis below in the discussion section. 

Cheema et al.\cite{Cheema2020} found that very thin films of only a couple unit cells thick have enhanced ferroelectric response and lattice parameters that differ significantly from thicker films or bulk, but we do not address such thin films here using bulk computations. Thicker films of 3-10 nm have lattice parameters consistent with our computations even at zero stress, consistent with previous bulk computations\cite{Park2013,Schroeder_FE2019} and what we report below. Ref.~\onlinecite{Zhou2022} considers the large strains in bulk computations and find that they help promote stability of the ferroelectric \oI~phase. Here we show that such large strains are not necessary for ferroelectric behavior, consistent with experimental observations. 

In the literature, hafnia has been considered an improper ferroelectric, and several parent structures have been proposed for \oI, including fluorite cubic\cite{Lee2020}, $Ccce$\cite{Delodovici2021}, $Pbcm$\cite{Aramberri2023}, and tetragonal\cite{Huan2014,Zhou2022}. For ferroelectric \oII, some correctly considered tetragonal \tphase~as the parent structure\cite{Huan2014,Qi2020}, which would be a proper soft-mode transition. Since polar $R3m$ or $R3$ has a higher symmetry than most phases except fluorite cubic and centrosymmetric $R\bar{3}m$\cite{Ouyang2023}, it must have either cubic or rhombohedral as parent. Ref.~\onlinecite{Lee2020} concluded that ferroelectricity in hafnia is improper, and is due to antipolar-polar coupling and nonlinear interactions with the primary phonon instability in cubic $X^{-}_{2}$. In a similar but different way, Zhou et al. showed that tensile uniaxial strain and antipolar-polar coupling leads to the ferroelectric \oI~and claimed that hafnia is an improper ferroelectric\cite{Zhou2022}. Using a different parent structure $Ccce$ (equivalently $Bbab$) Ref.~\onlinecite{Delodovici2021} claimed that three modes,  $\Gamma_{3-}$, $Y_{3+}$, and $Y_{2+}$, are coupled to generate the \oI~structure. In contrast, Ref.~\onlinecite{Aramberri2023}, using $Pbcm$ as the parent structure, stated correctly that ferroelectricity in hafnia is proper, but $Pbcm$ is a higher energy parent, and gives deeper double wells than we show below.

In this paper we clarify symmetry relationships among the different phases of hafnia and the phonon modes that connect them. We find that antiferroelectric $Pbcn$ has a zone center unstable mode $\Gamma^{-}_{2}$, whose distortion leads to \oI. We further find that the double-well potential resulting from this $\Gamma^{-}_{2}$ distortion is shallow, making $Pbcn$ key in the polarization switching of \oI~hafnia.  Having a correct model for hafnia's ferroelectric modes and double wells is important for learning to control its properties by doping, strain, and field, as well for proper understanding of its material physics. In order to optimize its switching behavior and design domain engineering and best epitaxial growth, a correct understanding of the ferroelectric instabilities is crucial.

\section*{Methods}
We performed symmetry analyses starting with the fluorite structure and the many potential structures from subgroups of the fluorite structure using {\sc isodistort}\cite{Isodistort1,Isodistort2} and the Bilbao Crystallographic Server\cite{BilbaoI,BilbaoII,BilbaoIII}. We performed first-principles calculations using {\sc{Quantum Espresso}} (QE) \cite{QE-2009,QE-2017,QE-exa} using Garrity-Bennett-Rabe-Vanderbilt (GBRV)\cite{GBRV2014} pseudopotentials with PBEsol \cite{PBEsol2008} exchange-correlation. We used different exchange-correlation such as PBE\cite{PBE1996} but we find that PBEsol gives a volume of baddeleyite well in agreement with experimental observation, namely 1.04\% compared to Ref.\onlinecite{Pathak2020}. The plane-wave expansion cutoff energy E$_\textbf{cutoff}$ was 544 eV, and the Brillouin zone was sampled using an 6$\times$6$\times$6 Monkhorst-Pack (MP) grid \cite{MP1976}. Convergence tests were performed for all parameters, and calculations of the depth of the $Pbcn$ double well using different MP grid (4$\times$4$\times$4 and 8$\times$8$\times$8) gave similar $\Delta{E} \simeq$24 meV/atom. Phonon frequencies were obtained using Density Functional Perturbation Theory (DFPT) \cite{Baroni2001} implemented in the QE/PH package\cite{QE-2009,QE-2017,QE-exa} and the phonon dispersion curves along the $\Gamma$--X--M--$\Gamma$ high symmetry line of the Brillouin zone (Fig.\ref{Fig3}) are an interpolation of the interatomic force constants (IFC) computed using 4$\times$4$\times$4 q-vector grid. 
The computed energy differences, the relaxed lattice lattice parameters of the various polymorphs, and their Raman spectra compare well with previous studies\cite{Aldo2022,Aldo2023}.

We parameterized the studied distortions using the displacive order parameter $Q$ as the mode amplitude\cite{Isodistort3}: 
\begin{equation}\label{EqQ}
Q = \sqrt{\sum_{i}{\textbf{u}_{i}}^{2}}
\end{equation}
where ${\textbf{u}}_{i}$ is the displacement of atom $i$ in Angstroms (\AA).

\section*{Results}
\subsection*{Group-subgroup relations of different hafnia polymorphs}
\subsubsection*{Distortion of fluorite cubic \cphase}
\begin{figure*}[!th]
\centering
\includegraphics[width=0.65\textwidth]{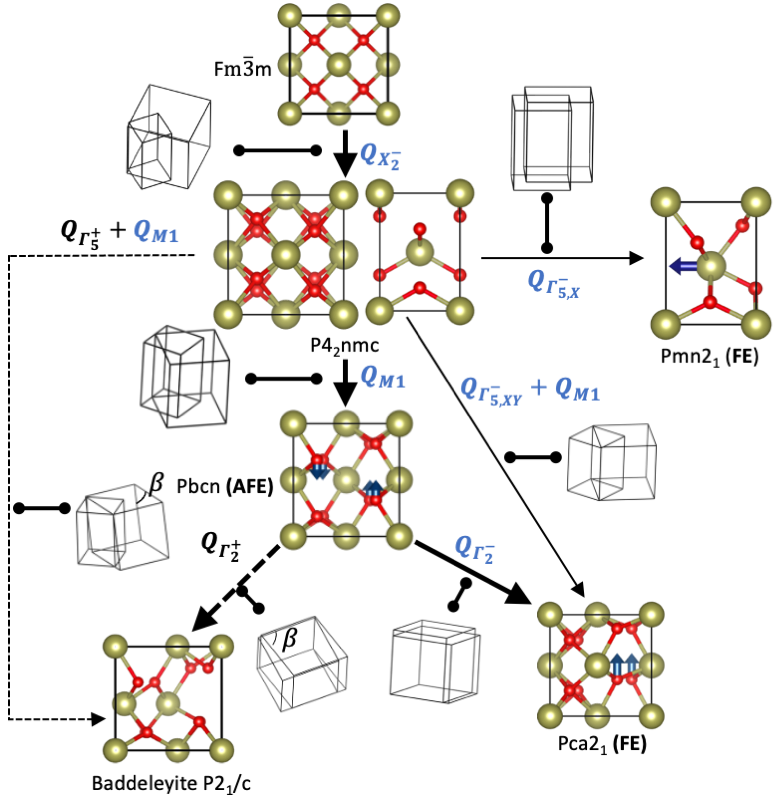}
\caption{Summary of the transitions to different hafnia polymorphs starting from cubic fluorite \cphase (topmost) to tetragonal \tphase, \oII~(top right), $Pbcn$ (middle), \oI~(bottom right), and baddeleyite (bottom left). The symmetry designations of the order parameter responsible for the distortion are labelled near the arrow, and are all written in terms of the parent phase (tail of the arrow), as it is done throughout this paper. The blue symmetry labels indicate soft modes and the outlines depict the unit cell relationships. Solid lines with arrows show soft-mode relationships, whereas dashed lines must be first-order transitions. The dark blue arrows on the atoms of the FE and antiferroelectric (AFE) structures show the direction of the polarization. The angle $\beta$ of the monoclinic structure is 99.06$^{\circ}$. Not all symmetries shown appear on the phase diagram, nor are all subgroups or pathways shown.}
\label{Fig1}
\end{figure*}

Firstly, we consider fluorite cubic (\cphase) because it is the aristotype for hafnia.  It has a three-dimensional soft mode  $X^{-}_{2}$ whose distortion can be denoted as $Q_{X^{-}_{2,i}}$, where $i={X,Y or Z}$ with a linear combination:
\begin{equation}
    Q_{X^{-}_{2,XYZ}} = a Q_{X^{-}_{2,X}} + b Q_{X^{-}_{2,Y}} + c Q_{X^{-}_{2,Z}}
\end{equation}
denoted by (a,b,c), where $a$,$b$, and $c$ are coefficients of the linear combination. The soft $X^{-}_{2}$ modes can lead to three possible distortions: \tphase~(Fig.\ref{Fig1}), $P4/nbm$, and $P\bar{4}3m$. The first two phases are tetragonal distortions of fluorite cubic according to (a,0,0), and (a,-a,0) respectively. There are possible pathways from $P4/nbm$ to $Pbcn$ or $Pbca$ but we do not explore these further because it has not been observed experimentally to the best of our knowledge. The third possible distortion of $X^{-}_{2}$ according to (a,a,a) is $P\bar{4}3m$, which is cubic, non-centrosymmetric and thus piezoelectric, but non-polar. Interestingly, $P\bar{4}3m$ has been reported in the literature as another possible cubic phase\cite{Barabash2017, Wang2004,Bichelmaier2023}. $P\bar{4}3m$ and \tphase~have similar powder X-ray diffraction (XRD) patterns with only small differences in the relative intensities. It can therefore hard to distinguish them in polycrystalline samples.

\begin{figure*}[!th]
\centering
\includegraphics[width=0.85\textwidth]{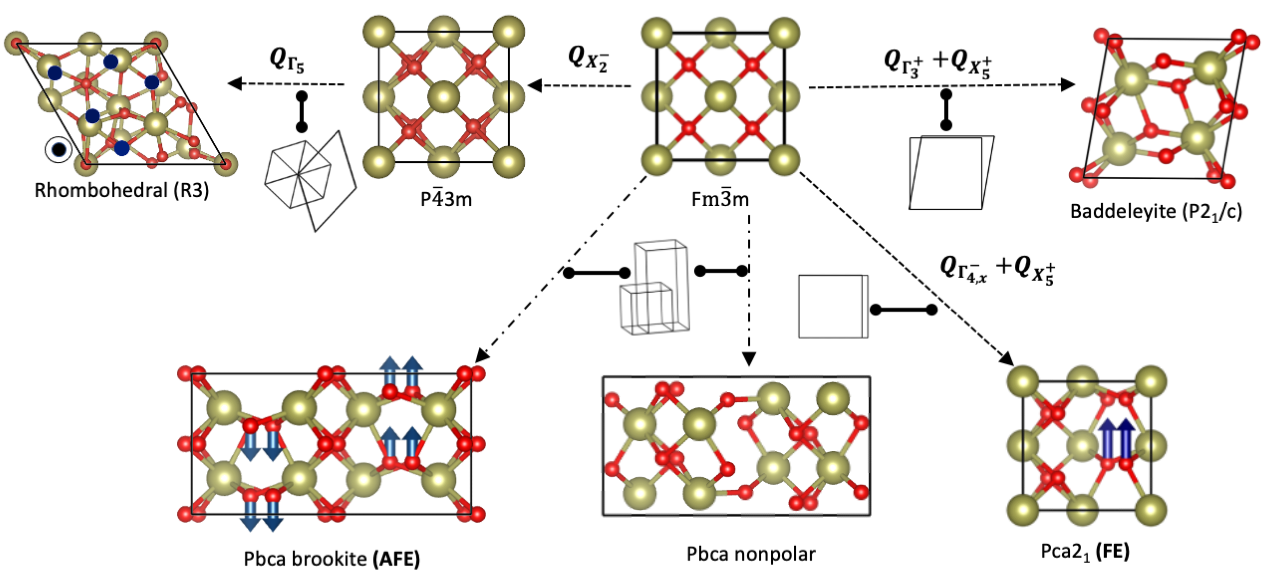}
\caption{Summary of the transitions to different hafnia polymorphs starting from cubic fluorite \cphase (center) to rhombohedral $R3m$ (left), \oI~(bottom right), orthorhombic $Pbca$ (bottom left), and baddeleyite (right). The symmetry designations of the order parameter responsible for the distortion are labelled near the arrow, and are all written in terms of the parent phase (tail of the arrow). The outlines depict the unit cell relationships. The polarization for the rhombohedral phase is out of the page as indicated, along the Cartesian \textbf{z}-direction. Possible soft-mode transitions are indicated by solid arrows, and possible first-order transition involving hard modes are all dashed. The dashed-dotted lines to the two $Pbca$ structures indicate either a first order transition from cubic to $Pbca$ or a complicated combination of 6 optical modes (cubic modes $X^{+}_{5}$, $X^{-}_{5}$, $X^{-}_{3}$, $W_{3}$, $W_{4}$, and $\Delta_{5} (0,\frac{1}{2},0)$) and strain. Both $Pbca$ structures (brookite antipolar and the nonpolar structure) have the same cubic modes involved in the coupling but their difference is in their linear combinations. Not all symmetries shown appear on the phase diagram, nor are all subgroups or pathways shown.}
\label{Fig2}
\end{figure*}

Two different $Pbca$ structures can be considered a distortion of \cphase~(Fig.\ref{Fig2}), and they are often confused. They were properly distinguished by Kersch and Falkowski, who called them nonpolar uP61 and antipolar aP61,  which is the brookite structure \cite{Kersch2021}. The designation ``OI'', as denoted in Ref.~\cite{Schroeder_FE2019}, is sometimes applied to brookite and sometimes to the nonpolar structure. They have the same Wyckoff positions, but different atomic coordinates (Table~\ref{Table1} and Fig.~\ref{Fig2}). The o-phase in Ref.\onlinecite{Materlik2015,Kersch2021} referred to the nonpolar structure, and in Ref.\onlinecite{Ohtaka1995,Kersch2021} to the antipolar brookite. Although brookite might be considered antiferroelectric (Fig.\ref{Fig2}), there is no phonon mode that connects it to \oI~nor is there a group-subgroup relationship between them.
Complicated coupling among six modes plus strain leads to $Pbca$ from cubic, which would be first-order or have intermediate structures (as in Ref.~\onlinecite{Nentwich2022}). Coupled distortions (mainly cubic modes $X^{+}_{5}$ and $X^{+}_{3}$) lead to baddeleyite. We stress that in cubic fluorite \cphase, the only soft mode is the zone boundary $X^{-}_{2}$. The other distortions of cubic, which lead to \oI, $Pbca$, and baddeleyite, are all hard modes.

\subsubsection*{Distortion of tetragonal \tphase~and $Pbcn$}
The high temperature tetragonal \tphase~phase is also a precursor to antiferroelectric and ferroelectric orthorhombic phases and the ground state baddeleyite. Using group theory we investigate how baddeleyite, \oI, \oII, and the $Pbcn$ can be obtained from a transformation of the tetragonal phase. 

We find that tetragonal \tphase~is related to antiferroelectric $Pbcn$ by a displacive order parameter with $M1$ symmetry (Fig.\ref{Fig1}). $M1$ is a soft mode of extensionally strained tetragonal \tphase (later subsections). \oI~is related to tetragonal \tphase~through the coupling of the $M1$ and $\Gamma^{-}_{5}$ modes, and baddeleyite results from the coupled $M1$ and $\Gamma^{+}_{5}$ distortions. We use simplified symmetry designations of the parent structure modes (tetragonal \tphase~in this case) according to {\sc isodistort}\cite{Isodistort1,Isodistort2}. We found no group-subgroup relation between \oI~and baddeleyite, so any transition from ground state baddeleyite to ferroelectric \oI~must be through nucleation and growth.

Although not observed experimentally, \oII~is the subject of many computational studies\cite{Qi2020,Huan2014,Barabash2017}. Tetragonal \tphase~is distorted into the \oII~structure by its $\Gamma^{-}_{5,X}$ mode (equivalently $\Gamma^{-}_{5,Y}$).

$Pbcn$ is a parent of \oI~and baddeleyite \mHfo. A distortion of $Pbcn$ with $\Gamma^{-}_{2}$ or $\Gamma^{+}_{2}$ symmetry generates \oI~and baddeleyite \mHfo~respectively (Fig.\ref{Fig1}). The $\Gamma^{-}_{2}$ mode is soft for relaxed $Pbcn$.

There are lots of local minima, which complicates the relationships among the metastable phases. Any of these metastable and saddle point structures might be stabilized entropically with temperature by anharmonic thermal vibrations or hopping.

\subsubsection*{Distortion of cubic $P\bar{4}3m$}
Ferroelectricity in hafnia films on hexagonal substrates is also be attributed to polar rhombohedral phases like $R3m$ or $R3$\cite{Noheda_Jorge2020}.
Symmetry investigations indicates that $P\bar{4}3m$ can be distorted into $R3m$ and $R3$ with a single polar mode $\Gamma_{4}$ or $\Gamma_{5}$ respectively, but $P\bar{4}3m$ is lower in total energy than cubic fluorite, as well as the rhombohedral phases ($R3m$, $R3$, and $R\bar{3}m$) for a relaxed lattice. The dynamical instability of the relaxed $P\bar{4}3m$ is a two-dimensional zone center soft mode that leads to tetragonal $P\bar{4}2m$ and orthorhombic $P222$. Nonetheless, there are conditions under which $P\bar{4}3m$ can become the parent of the polar rhombohedral phases. Under tensile strain (e.g. $a$=5.26 \AA, where $a$ is the cubic lattice constant), a three-dimensional zone-center instability ($T1$) develops in addition to the previously mentioned $E$ mode. One of the $T1$ modes distorts strained $P\bar{4}3m$ into $R3m$, which is more stable by about 6 meV/atom. This indicates that ferroelectricity in rhombohedral hafnia is also proper with a $P\bar{4}3m$ parent structure. The conclusion of improper ferroelectricity in polar rhombohedral hafnia by Ref.\onlinecite{Ouyang2023} is likely because of the use of $R\bar{3}m$ as the parent structure, which is higher in energy than $P\bar{4}3m$.

\begin{center}
\begin{table}[h!]
\centering
\begin{adjustbox}{width=0.65\columnwidth,center}
\begin{tabular}{|c|c|c|c|c|}
\hline
& &\textbf{a} & \textbf{b}  & \textbf{c} \\
\hline
\multirow{4}{*}{Pbca (oI) } & &  5.13 \AA & 5.25 \AA  & 10.07 \AA \\
& Hf & 0.042  & 0.157 & 0.862  \\ 
& O  & 0.169  & 0.666 & 0.466 \\
& O  & 0.253  & 0.589 & 0.224 \\
\hline
\multirow{4}{*}{Pbca\cite{Materlik2015}} & & 5.15 \AA & 5.29 \AA  & 10.14 \AA \\
& Hf & 0.041 & 0.343  & 0.138 \\
& O  & 0.325 & 0.160  & 0.032 \\
& O  & 0.248 & 0.589  & 0.275 \\
\hline
\multirow{4}{*}{Brookite} & & 5.03 \AA &  9.95 \AA & 5.20 \AA \\
& Hf & 0.246 & 0.116  &  0.536 \\
& O  &  0.002 & 0.023  &  0.239 \\
& O  & 0.128  & 0.710  &  0.875 \\
\hline
\multirow{4}{*}{Pbca\cite{Ohtaka1995}} & & 5.06 \AA & 10.02 \AA & 5.23 \AA \\
& Hf & 0.244 & 0.385  & 0.034 \\
& O  & 0.372 & 0.290  & 0.372  \\
& O  & 0.000 & 0.023  & 0.243 \\
\hline
\end{tabular}
\end{adjustbox}
\caption{Lattice parameters and atomic positions of different Pbca structures. \textbf{a}, \textbf{b}, and \textbf{c} are the crystallographic direction of the orthorhombic Bravais Lattice. The atomic positions are given relative to the Bravais lattice vectors. All the atoms sit on  Wyckoff position 8c.}
\label{Table1}
\end{table}
\end{center}

\subsection*{Phonon dispersion}
\begin{figure}[!ht]
\centering
\includegraphics[width=0.75\columnwidth]{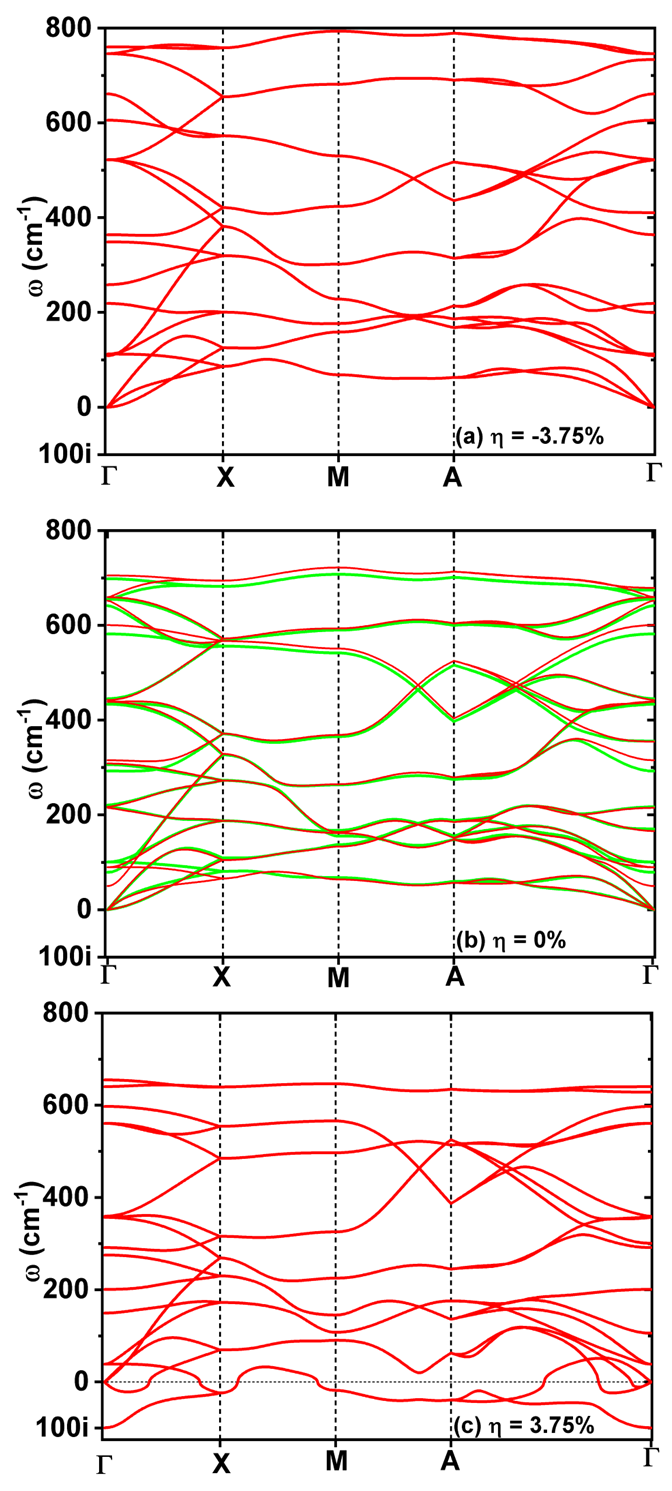}
\caption{Phonon dispersion of tetragonal \tphase~at strain (a) $\eta=$ -3.75 \%, (b) $\eta =$0\%, and (c) $\eta =$3.75\%. The dotted lines on panel (c) shows the zero frequency level. The epitaxial strain $\eta$ is a biaxial in-plane strain with $c$ allowed to relax, so that $a_{\eta} =  (1+\eta) a_{0}$
where $a_{0}$ is the lattice constant the stress-free tetragonal ($a_{0}$ =5.00\AA). Shown are the harmonic frequencies, i.e. the square root of the curvature of the potential surface for atoms in their ideal positions. The red and green curves in panel (b) are the phonon dispersion curves obtained using PBEsol and PBE exchange correlation respectively.}
\label{Fig3}
\end{figure}

The tetragonal phase is dynamically stable at zero strain and under compressive strain ($ 0 \leq \eta \leq$ -3.75\%) (Fig.\ref{Fig3}(a) and (b)). However, under tensile strain ($\eta \geq 2.0\%$ or \aep $\geq$ 5.13\AA) tetragonal hafnia becomes unstable. Indeed for $\eta = 3.75\%$, corresponding to \aep = 5.22 \AA~(Fig.\ref{Fig3}(c)), we find soft modes at the Brillouin zone center ($\Gamma^{-}_{5}$), the zone boundary points M ($M1$), X, and A. The lowest branch is soft along the whole M--A--$\Gamma$ line, indicating the possibility of incommensurate or disordered ferroelectric/antiferroelectric ordering possibly explaining previous observations\cite{Lee2020}.
Along the A--$\Gamma$ portion of the phonon dispersion curve around $\omega =$ 400 cm$^{-1} $, we observe avoided crossing of the optical phonon branches. This involves optical branches of the same symmetry, which tracks back to the $E_{u}$ at the $\Gamma$ point. The frequency difference between these two branches decreases as the tensile strain increases, i.e.\ the strain becomes more positive.

We calculated the phonon dispersion of the relaxed tetragonal using PBE and PBEsol (Fig.\ref{Fig3}b), and they are in excellent agreement. Calculation of the dispersion curves within the local density approxiamtion (LDA)\cite{PZ1981} also confirmed the dynamical stability of relaxed \tphase. Small quantitative differences are nevertheless found, for example in the onset of the dynamical instability in strained tetragonal. The first soft mode $\Gamma^{-}_{5}$ is observed around 1.5\% tensile strain (a$_{epi}$ = 5.05 \AA) when using LDA, but around 2.0\% tensile strain (a$_{epi}$ = 5.13 \AA) with PBEsol.

\subsection*{Soft phonon eigendisplacements}
The symmetry mode for the double well in the ferroelectric \oI~phase has $\Gamma_{2}^{-}$ symmetry in the antiferroelectric parent $Pbcn$. It is an unusual case, though, compared with what one is used to seeing for polar modes in perovskite ferroelectrics. At high temperatures the atoms can hop between the up and down positions in $Pbcn$, which is the tetragonal \tphase~$M1$ mode, leading to the tetragonal structure.

\begin{figure}[!ht]
\centering
\includegraphics[width=0.45\columnwidth]{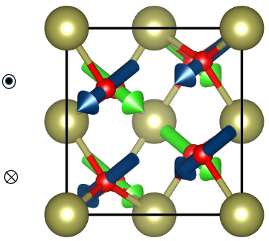}
\caption{The arrows show the polar, soft $\Gamma_{2}^{-}$ mode of $Pbcn$ that leads to \oI. The $Pbcn$ antipolar direction is top to bottom, and the blue and green arrows indicate displacements of front and back oxygen atoms respectively. The top oxygen atoms are moving out of the page and the bottom ones are moving into the page as indicated by the symbols adjacent to the structure. \oI~can be thought of as polar in one direction ($z$), and antipolar in the perpendicular $x$ and $y$ directions. The displacements of the hafnium atoms are small in comparison to the oxygen atoms, so are not shown.}
\label{Fig4}
\end{figure}

\begin{figure}[!ht]
\centering
\includegraphics[width=\columnwidth]{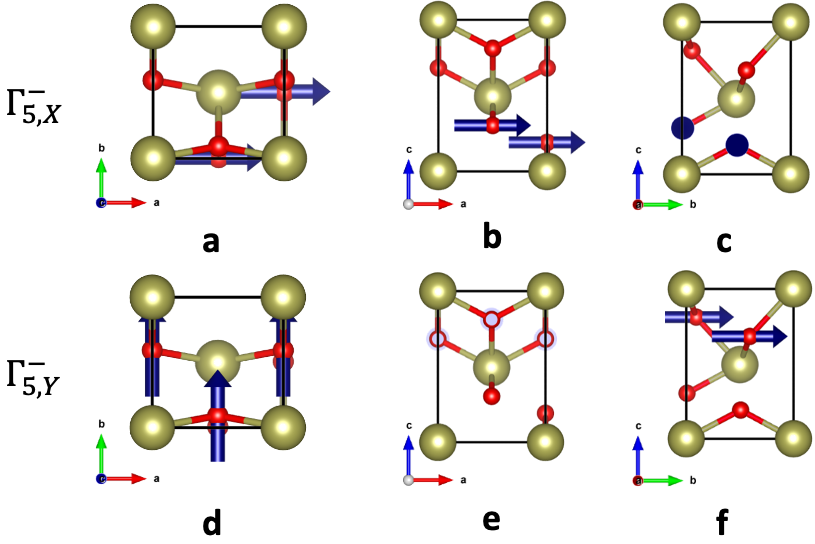}
\caption{The doubly degenerate $\Gamma_{5}^{-}$ soft modes of tetragonal \tphase, denoted by $\Gamma^{-}_{5,X}$ (panels a, b , and c); and $\Gamma^{-}_{5,Y}$ (panels d, e, and f). The structures were projected on the ab plane (panels a and d), ac plane (panels b and e), and bc plane (panels c and f). The blue arrows on the atoms indicate the directions of their displacements.}
\label{Fig5}
\end{figure}

The $\Gamma^{-}_{5}$ mode of tetragonal \tphase~becomes unstable at 2.0\% epitaxial strain (\aep = 5.13 \AA). It is a two-dimensional mode, with two symmetrically equivalent displacive order parameters, denoted $Q_{\Gamma^{-}_{5,X}}$ and $Q_{\Gamma^{-}_{5,Y}}$, represented as (a,0) and (0,a), respectively using {\sc{isodistort}}\cite{Isodistort1,Isodistort2} notation. We find that the eigendisplacements of the $\Gamma^{-}_{5}$ modes are related by screw symmetry ($4_{2}$), i.e.\ ~the combination of 90$^{\circ}$ rotation about the \textbf{c}-axis followed by a translation by half the lattice constant along the \textbf{c} (Fig.\ref{Fig5}). The distortion of either $Q_{\Gamma^{-}_{5,X}}$, or $Q_{\Gamma^{-}_{5,Y}}$ by itself leads to the \oII~ ferroelectric phase, and its linear combination:
\begin{equation}
     Q_{\Gamma^{-}_{5,XY}} = a~Q_{\Gamma^{-}_{5,X}} +  b~Q_{\Gamma^{-}_{5,Y}},
\end{equation} 
distorts tetragonal \tphase~into orthorhombic $Aba2$ when the coefficients $a = b$. $Aba2$ is a saddle point and relaxes to centrosymmetric $Bbab$. $Bbab$ is stable and is 29 meV/atom above baddeleyite in energy.

\begin{figure}[!ht]
\centering
\includegraphics[width=\columnwidth]{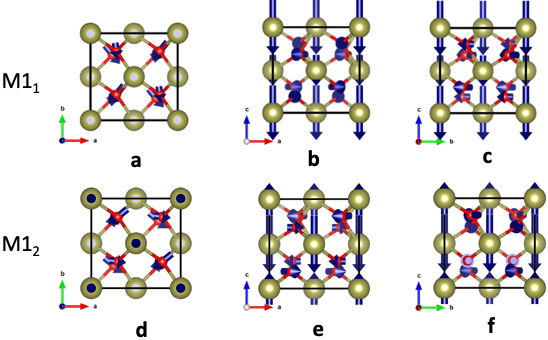}
\caption{The doubly degenerate soft $M1$ modes of tetragonal \tphase, denoted by $M1_{1}$ (panels a, b , and c); and $M1_{2}$ (panels d, e, and f). The structures were projected on the ab plane (panels a and d), ac plane (panels b and e), and bc plane (panels c and f). The blue arrows on the atoms indicate the directions of their displacements.}
\label{Fig6}
\end{figure}

$Pbcn$, the centrosymmetric parent of ferroelectric \oI, is generated from tetragonal \tphase~from the $M1$ distortion (Fig.\ref{Fig1}). The order parameter associated with the doubly degenerate soft $M1$ mode is denoted by $Q_{M1}$ and represented as (a,0), and their eigendisplacements (Fig.\ref{Fig6}) are related by $4_{2}$ screw symmetry. 

\begin{figure}[!ht]
\centering
\includegraphics[width=\columnwidth]{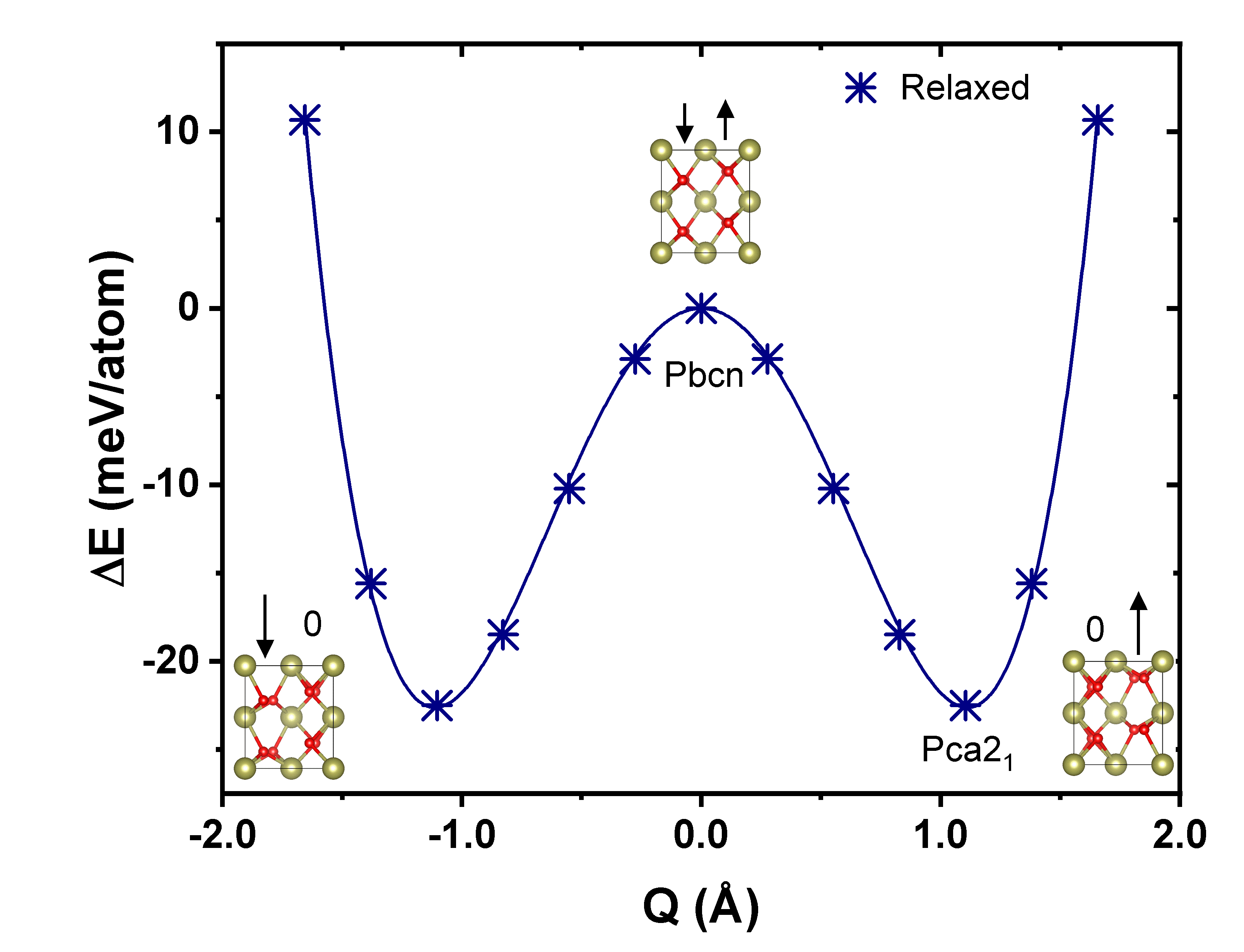}
\caption{Potential well for the $\Gamma^{-}_{2}$ symmetry mode leading from $Pbcn$ to \oI for relaxed lattice. The structures corresponding to the three extrema of the double well are drawn above them, and the polarization switching is shown by the polarization configurations: $\textbf{0} \uparrow$ (right well), $\downarrow \uparrow$ (top of double well), and $\downarrow \textbf{0}$ (left well).}
\label{Fig7}
\end{figure}

In $Pbcn$ the oxygen displacements can be considered as up and down along the antipolar axis (Fig.\ref{Fig7}) and the polar mode (Fig.\ref{Fig4}) brings one of these displacements (or local polarizations) for half the oxygen atoms in the unit cell to zero, relative to the surrounding hafnia atoms, while the displacement for the other set of oxygens grows to double that in the parent $Pbcn$ phase. When the polarization is reversed, the displaced oxygens move to the center of their polyhedra, while the other set of oxygens gain a large polarization. So the parent phase has ``up, down''  polarization, where one side of the double well has ``zero, double down'' and the other side has ``double up, zero''. The parent structure can be considered as an ordered antiferroelectric.

For completeness, we also discuss the stabilization of ferroelectric \oII. For an extensional epitaxial strain $\eta \geq 2.0\%$ (\aep $\geq$ 5.13\AA), \oII~is more stable than \tphase~along $Q_{\Gamma^{-}_{5,X}}$.

\begin{figure}[!ht]
\centering
\includegraphics[width=\columnwidth]{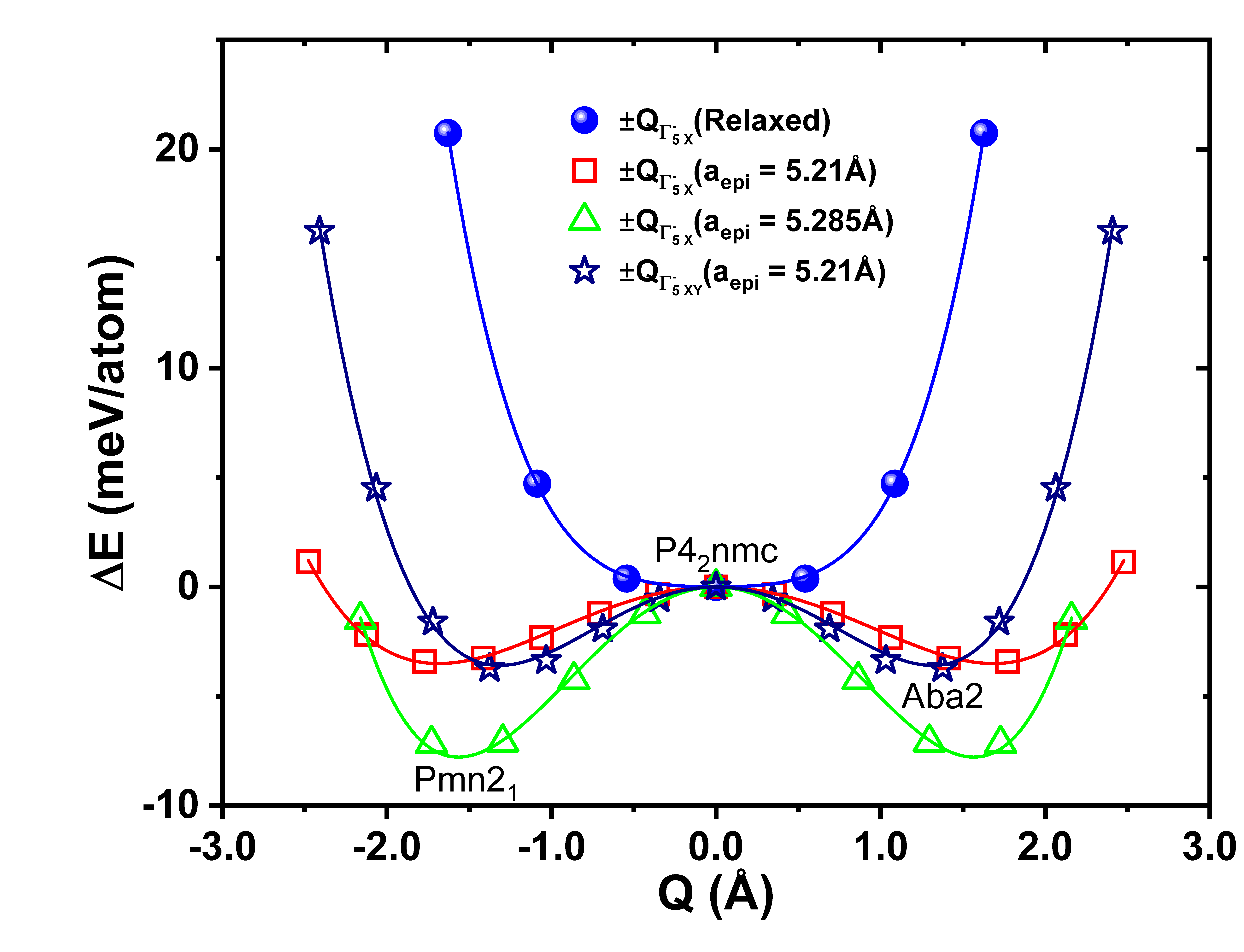}
\caption{Energy as a function of $\Gamma^{-}_{5,X}$ and  $\Gamma^{-}_{5,XY}$ distortions for relaxed tetragonal \tphase (blue curve), and for \aep = 5.21 \AA, and at \aep = 5.285 \AA.}
\label{Fig8}
\end{figure}

\begin{figure}[!ht]
\centering
\includegraphics[width=\columnwidth]{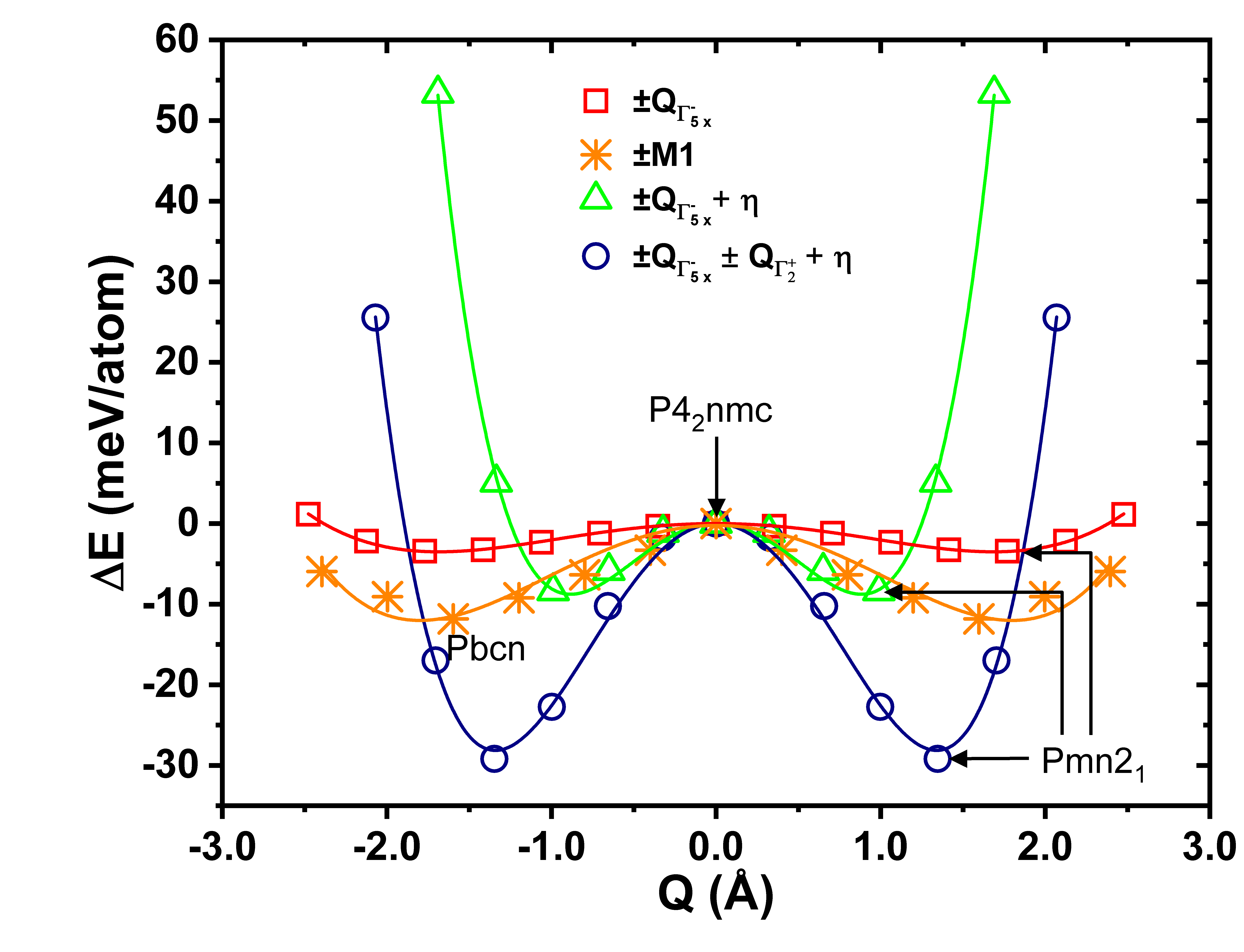}
\caption{Energy as a function of distortion for \oII. The curves are labelled according to the irreducible representation, and $\eta$ indicates the tetragonal lattice parameters are relaxed. The symmetry of the parent and distorted structure are labelled near the top and bottom of the double-well, respectively. The red, green and blue curves were obtained for \aep = 5.21 \AA, whereas the orange curve was obtained for \aep = 5.25 \AA.}
\label{Fig9}
\end{figure}

\subsection*{Discussion}
We have shown that there are single polar soft modes that distort the tetragonal and the $Pbcn$ parent phases to the polar \oII~and \oI~phases, respectively, so that hafnia is a proper ferroelectric. The centrosymmetric parent for the experimentally observed \oI, is $Pbcn$, which is dynamically unstable. 

We now consider possible ways to access $Pbcn$ and thus \oI, since they are not known to be stable at any conditions of pressure and temperature. Hafnia-based samples are usually annealed at high temperature ($>$773K)\cite{Hongrae2022}.
Although, the temperature--pressure phase diagram of pure hafnia shows that tetragonal is the stable phase at ambient pressure within the temperature range of 1800K to 2600K\cite{Ohtaka2001}, the transition temperature usually decreases with doping concentration\cite{Hoffmann2015,ParkSchenk2017} so that the high temperatures ($>$773K) used in experiments\cite{Hongrae2022} in doped hafnia are possibly large enough to stabilize the tetragonal phase. The strain imposed by the substrate (e.g.\ yttria-stabilized zirconia) destabilizes the tetragonal, which distorts into $Pbcn$.

Although Ref.~\onlinecite{Zhou2022} indicated $Pbcn$ as a key intermediate phase leading to \oI, the parent considered was the tetragonal phase which led to their conclusion about the improper nature of ferroelectricity in hafnia.

In considering $Pbcm$ as the reference structure instead of cubic, Ref.~\onlinecite{Aramberri2023} showed the existence of a single soft polar mode that gives \oI, also consistent with proper ferroelectricity in hafnia. $Pbcm$, however, is not a subgroup of the tetragonal phase. The double well depth from $Pbcn$ to \oI~is shallower than $Pbcm$, (~72 meV/f.u vs 117 meV/f.u.\cite{Aramberri2023}). The correct switching pathway is the lowest energy pathway. 

The lowest energy parent structure we find for the ferroelectric \oI~phase is $Pbcn$, which is thus the low energy pathway for switching the polarization direction in hafnia. There may not be a thermodynamically stable phase field for $Pbcn$--that is not required. However, even if formed metastably, it would probably consist of disordered off-centered cells with atoms hopping back and forth between double wells, and have at least partly order-disorder character.

The B\"{a}rnighausen tree of the Hf$_{x}$Zr$_{1-x}$O$_{2}$ solid solution\cite{Nentwich2022} gives the group-subgroup relations between different hafnia polymorphs. The phases and paths shown there are not complete, since the parent of different structures and the sequence of transitions depends on which mode is first frozen. The fact that B\"{a}rnighausen trees are not unique may not have been generally recognized. For example in the cubic to $Pbca$ transition, the freezing of the $W3$ or $X^{-}_{3}$ distorts the cubic into $I4_{1}/acd$ or $P4/nmm$, which are possible distortions not shown in Ref.\cite{Nentwich2022}. On the other hand, the number of subgroup possible is infinite as pointed out by B\"{a}rnighausen\cite{Barnighausen1980} so there is always a choice of which to include. Indeed other possible cubic phases of hafnia can exist such as $P\bar{4}3m$\cite{Barabash2017,Bichelmaier2023}. The latter phase can also be considered a distortion of the aristotype fluorite cubic ($Fm\bar{3}m$). The goal here is to concentrate on low energy structures and the most probable routes to them from the high temperature cubic and tetragonal phases. We focus on distortions involving optical phonons, and provide physical bases of phase transition through DFT computations. Additionally, the soft phonon modes and the ensuing double well potential of the $Pbcn$ indicates that there is no need for reconstructive transitions as proposed by Ref.~\onlinecite{Nentwich2022}. We find the maximum ion displacements from $Pbcn$ to baddeleyite or \oI~are only 0.45 \AA~and 0.25 \AA~respectively, much smaller than shown in Ref.~\onlinecite{Nentwich2022}.

We do not consider disordered phases or static structures with partially occupied sites such as included in Ref.~\onlinecite{Barabash2017,Nentwich2022} to be likely as static disordered structures are rare in non-molecular crystals. Common examples in molecular crystals are CO and H$_2$O ice Ih \cite{Clayton1932,Pauling1935}. Disorder in ferroelectrics is more likely to be of a hopping nature, such as in BaTiO$_{3}$ and KNbO$_{3}$\cite{Comes1970}, where in the 8-site model the atoms hop to give an average cubic structure.  $Pbcm$ included in Ref.\onlinecite{Barabash2017,Nentwich2022} as a parent of $Pbca$ and \oI was refined with half-occupied sites from {\em in situ} diamond anvil cell X-ray data for ZrO$_2$ \cite{Kudoh1986}, but neutron diffraction on quenched samples showed ordered $Pbca$ \cite{Ohtaka1990,Haines1997}. 

\section*{Conclusions}
The number of possible phases, local minima, and saddle points for the ABO$_2$ fluorite aristotype are truly astounding. We have shown the symmetry relations between the high temperature phases of hafnia to the relevant low symmetry structures, baddeleyite, \oI, \oII, AO $Pbca$, $R3m$ and the $Pbcn$. Tetragonal \tphase~hafnia is dynamically unstable under tensile strain, and the soft mode $M1$ leads to the antiferroelectric $Pbcn$ structure. $Pbcn$ is the best centrosymmetric parent for ferroelectric \oI. Ferroelectricity in hafnia is proper and follows the conventional soft mode theory. The shallow depth of the $Pbcn$ soft mode potential well is responsible for polarization switching under applied electric field.

\begin{acknowledgments}
A.R. and R.E.C thank P. Zubko, Hugo Aramberri and Jorge \'{I}\~{n}iguez for helpful discussions. This work is supported by U. S. Office of Naval Research Grant N00014-20-1-2699, and the Carnegie Institution for Science. Computations were supported by  high-performance computer time and resources from the DoD High Performance Computing Modernization Program, Carnegie computational resources, and REC gratefully acknowledges the Gauss Centre for Supercomputing e.V. (https://www.gauss-centre.eu/) for funding this project by providing computing time on the GCS Supercomputer SuperMUC-NG at Leibniz Supercomputing Centre (LRZ, www.lrz.de). 
\end{acknowledgments}

\bibliography{Hafnia}

\end{document}